\documentclass[12pt]{iopart}
\usepackage[pdftex]{graphicx}

\begin{document}

\title[Efficient basis for the Dicke Model I]{Efficient basis for the Dicke Model I: theory and convergence in energy}

\author{Miguel A. Bastarrachea-Magnani and Jorge G. Hirsch}

\address{Instituto de Ciencias Nucleares,
Universidad Nacional Aut\'onoma de M\'exico, Apdo. Postal 70-543, Mexico D. F., C.P. 04510}
\ead{miguel.bastarrachea@nucleares.unam.mx}
\begin{abstract}
An extended bosonic coherent basis has been shown by Chen {\it et al} \cite{Chen08} to provide numerically exact solutions of the finite-size Dicke model. The advantages in employing this basis, as compared with the photon number (Fock) basis, 
are exhibited to be valid for a large region of the Hamiltonian parameter space by analyzing the converged values of the  ground state energy.
\end{abstract}


\section{Introduction}
The Dicke Hamiltonian describes a system of $\mathcal{N}$ two-level atoms interacting with a single monochromatic electromagnetic radiation mode within a cavity \cite{Dicke54}. The Hamiltonian is very simple but not exactly solvable, and continues to drive research into its properties. One of the most interesting is its quantum phase transition (QPT) in the thermodynamic limit \cite{Hepp73,Wang73}. The interest on solving the Dicke Hamiltonian for a finite $\mathcal{N}$ comes not only from the fact that it provides a good description for the systems manipulated in the laboratory, but for the close connection found between quantum phase transitions,  entanglement, and quantum chaos \cite{Emary03,Lam05,Vid06}. Further, Dicke-like Hamiltonians have attracted much attention because of the experimental realization of the superradiant phase transition in a BEC \cite{Bau10,Nag10}.

The Dicke Hamiltonian is integrable for a finite $\mathcal{N}$ in two limits: when the atom-field interaction or the atomic energy gap are zero. In both situations, the atomic sector is described in the angular momentum basis, which is finite with $j=\mathcal{N}/2$. The no interaction case is diagonal in the photon number (Fock) basis, while the generalized bosonic coherent basis allows to construct analytic eigenstates of the Hamiltonian when the atomic excitation energy goes to zero \cite{Chen08,Basta11}. Employing the later a reduction by orders of magnitude in the size of the truncated subspace, which allows to obtain converged values of the observables, is found  \cite{Chen08,Basta11}. 
 
 The purpose of this work, and the accompanying one \cite{Hir13}, is to show that the benefits to employ the coherent basis are valid for a large region of the Hamiltonian parameter space, for the ground state and for a significative part of the energy spectra. In this first part the convergence in the energy is the criteria of choice.


\section{The Dicke Hamiltonian and its Integrable Limits}

The interaction between a system of $\mathcal{N}$ two-level atoms and a single mode of a radiation field can be described by the Dicke Hamiltonian: 
\begin{equation}
H_{D}=\omega a^{\dagger}a + \omega_{0} J'_{z} + \frac{\gamma}{\sqrt{\mathcal{N}}}\left(a+a^{\dagger}\right)\left(J'_{+}+J'_{-}\right).
\end{equation}
The frequency of the radiation mode is $\omega$, which has an associated number operator $a^{\dagger}a$. For the atomic part $\omega_{0}$ is the excitation energy, meanwhile $J'_{z}$, $J'_{+}$, $J'_{-}$, are collective atomic pseudo-spin operators which obey the SU(2) algebra.
The subspace of interest is defined by $j=\mathcal{N}/2$. In the thermodynamic limit the QPT takes place when the interaction parameter $\gamma$ reaches the critical value $\gamma_{c}=\sqrt{\omega\omega_{0}}/2$. 
At zero interactions the eigenstates are the tensor product between photon number, Fock states $|n\rangle$ for the radiation modes and angular momentum eigenstates $|j,m'\rangle$ for the atomic part, which we call the {\it Fock basis}. 


When the atomic frequency goes to zero, we have another integrable limit. It is obtained performing
a $-\pi/2$ rotation of the pseudospin operators around the y-axis $ J'_{z}=-J_{x}\,\,\mbox{,}\,\,\,J'_{x}=J_{z} \,\,\mbox{with}\,\,\,J_{x}= \frac{J_+ + J_-}{2}, $ and shifting the bosonic annihilation operator by 
$A=a+\frac{2\gamma}{\omega\sqrt{\mathcal{N}}}J_{z}=a+GJ_{z}.$ Substituting both transformations in the Hamiltonian we obtain:
\begin{equation}
H_{D}=\omega\left(A^{\dagger}A-G^{2}J_{z}^{2}\right)-\frac{\omega_{0}}{2}\left(J_{+}+J_{-}\right). 
\end{equation}
When $\omega_{0}\rightarrow0$ the Dicke Hamiltonian eigenstates are the tensor product between the number eigenstates $|N\rangle$  of $A^{\dagger} A$ and the angular momentum eigenstates $|j,m\rangle$ of $J_{z}$ \cite{Chen08,Basta11}. The vacuum of $A$ is an eigenstate of the annihilation operator $a$ with eigenvalue $\alpha=-Gm$. It is a coherent state seen in the Fock basis and the ground state of $H_{D}$ when $\omega_{0}\rightarrow 0$. We call them the {\em coherent basis}. In what follow we associate this basis with the use of a capital $N$, while for the Fock basis we employ the small $n$.

The coherent states depend on the angular projection eigenvalue $m$ of the atomic state, and on the interaction parameter $\gamma$ ($G = 2\gamma/\omega\sqrt{\mathcal{N}}$\,). When the interaction becomes null, the exact solution in the zero interactions limit is recovered. It implies that the coherent basis contains as a particular case the Fock basis in the situation in which it is the exact solution. 

In the next section we describe the numerical diagonalization.


\section{Numerical Diagonalization}

While the Fock basis is commonly used to diagonalize the Hamiltonian it becomes
intractable in the strong coupling limit even for a few dozens of atoms \cite{Chen08}. 

We compare the minimal truncation needed to obtain convergence of the solution, using the Fock and the coherent basis to diagonalize the Hamiltonian, truncating the bosonic sector up to $n_{max}$ and $N_{max}$, with energies $E^k_F$ and $E^k_C$ for the ground state, respectively. The convergence criteria is defined by using the upper limits $n_{max}$ and $n_{max}+1$, with some tolerance $\epsilon$.
\begin{equation}
\Delta E_{F}=|E^{G.S.}_F(n_{max}+1)-E^{G.S.}_F(n_{max})|<\epsilon .
\label{DeltaEF}
\end{equation} 
We consider the solution {\it converged} if the above relation holds. The same goes for the coherent basis. The same criteria will be used to find the minimum values $n_{max}$ and $N_{max}$ for any excited level $k$. In the next section we compare $n_{max}$ with $N_{max}$  in order to analyze the truncation behavior and the advantages of each basis in different parameter regions. 


\subsection{Ground State Energy}

\begin{figure}[h!]
\centering
\includegraphics[scale=0.7]{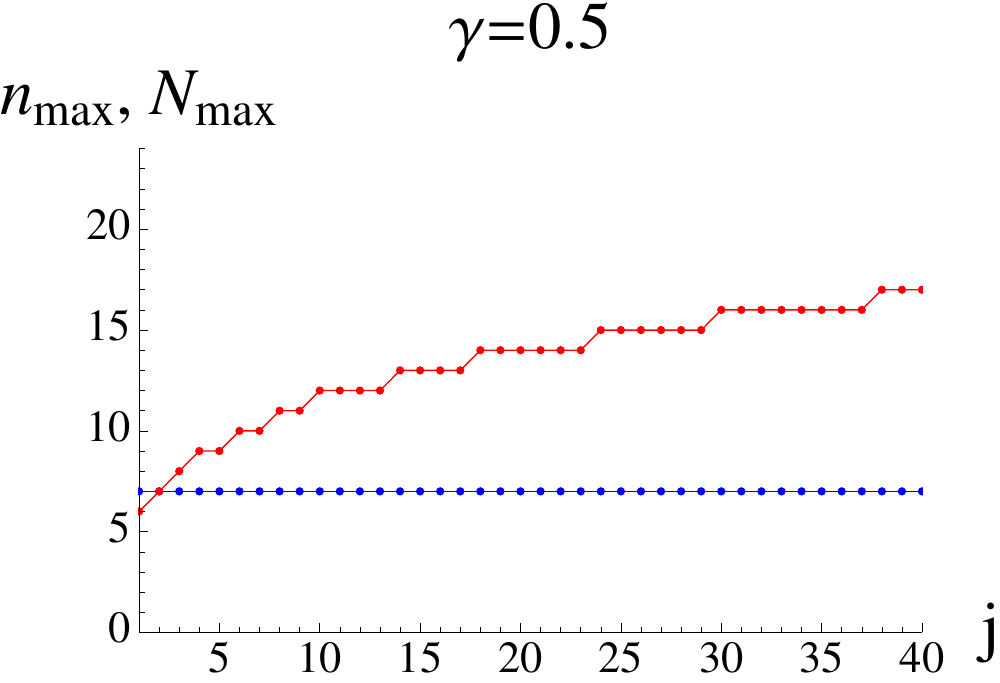}
\includegraphics[scale=0.7]{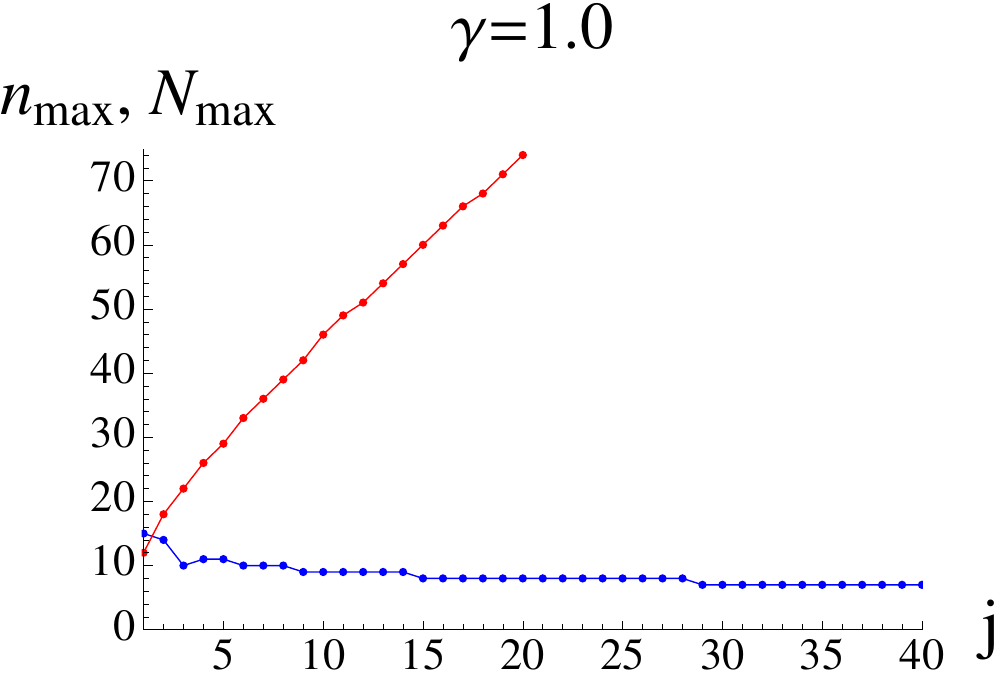}
\caption{$n_{max}$ (upper, red line) and $N_{max}$ (lower, blue line) as functions of $j$ with $\omega_{0}=\omega=1$, for $\gamma=0.5$ (left) and $\gamma=1.0$ (right). Tolerance $\epsilon=1$x$10^{-6}$.}
\label{fig1}
\end{figure}

In what follows we explore the truncation behavior for both basis in several parameter regions, principally for $\gamma$ and $\omega_{0}$. A comparison between the truncation in both basis is shown, in the terms explained above, for different values of the interaction parameter $\gamma$ and $j$, in resonance $\omega_{0}=\omega=1$, for the ground state. In this case the critical value of the interaction parameter is $\gamma_{c}=0.5$. The advantage of using the coherent basis becomes clear when the truncation is analyzed as a function of $j=\mathcal{N}/2$, from 1 to 40. As the atomic number $\mathcal{N}$ increases, the minimum number of photons needed to obtain convergence in the strong coupling limit grows too, making the numerical diagonalization very difficult using the Fock basis. For two representative interaction parameters, $\gamma=0.5$ and $1.0$, we show the results in figure \ref{fig1}. 

For the ultra-strong coupling $\gamma=1.0$ it is very difficult to obtain the numerical solution using the Fock basis for $j>20$ ($\mathcal{N}>40$) because it requires extremely large computing resources. This is the major problem of using the Fock basis. On the other hand, for the coherent basis, the dimensionality necessary for convergence is small and decreases as $j$ increases. 

\begin{figure}[tbp] 
\centering
\qquad
\includegraphics[width=0.45 \textwidth]{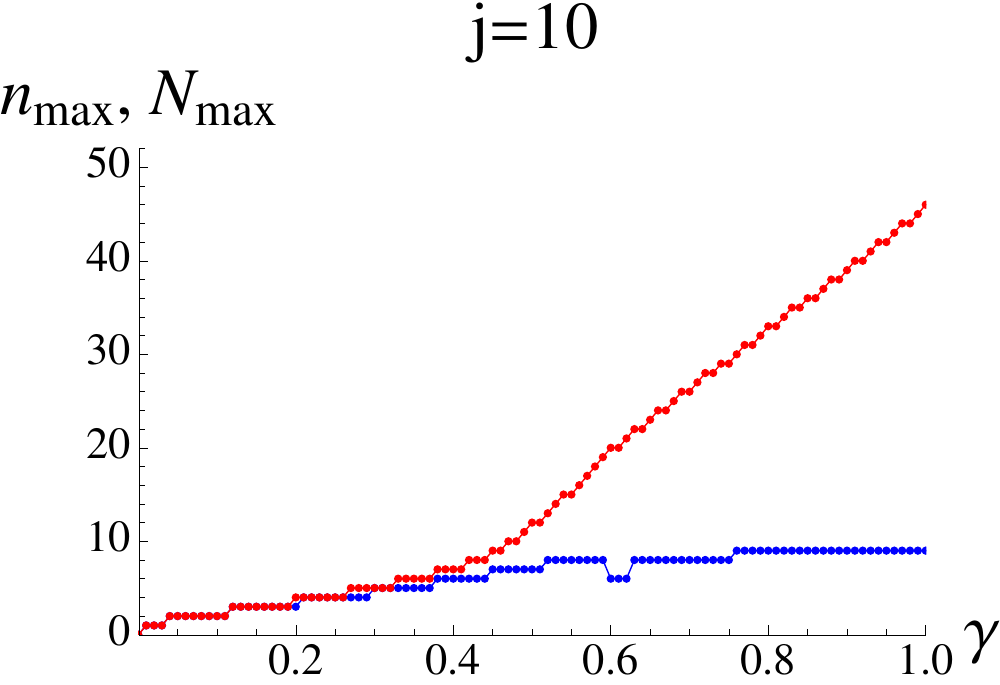}
\includegraphics[width=0.45 \textwidth]{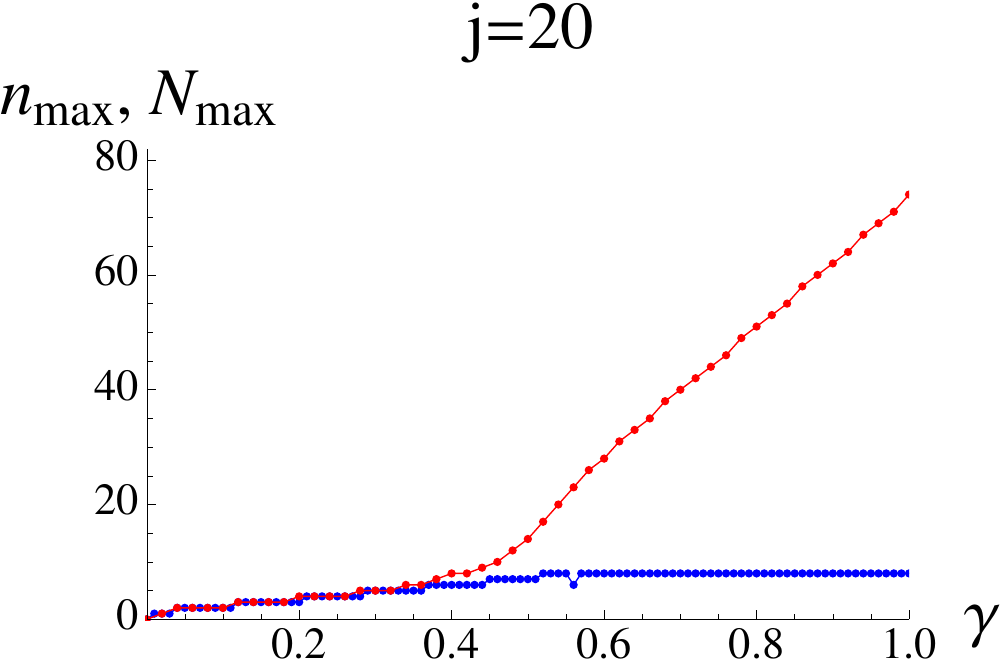}
\caption{$n_{max}$ (upper, red line) and $N_{max}$ (lower, blue line) as functions of $\gamma$ with $\omega_{0}=\omega=1$, for $j=10$ (left) and $j=20$ (right). Tolerance $\epsilon=1$x$10^{-6}$.}
\label{fig2}
\end{figure}

In figure \ref{fig2} we show the behavior of $N_{max}$ and $n_{max}$ as functions of $\gamma$, in resonance, for several representative values of $j$. Again, $\gamma_{c}=0.5$.

As we can see, meanwhile the value of $N_{max}$ increases slowly with $\gamma$, $n_{max}$ increases rapidly, making very hard the numerical diagonalization with the Fock basis for $j > 20$ and $\gamma \approx 1$. 
While the truncation in both basis is almost the same in the normal phase $\gamma < \gamma_c = 0.5$,  in the superradiant phase $n_{max}$ increases noticeably faster than $N_{max}$. 

Employing projected atomic SU(2) coherent states, an analytical expression for the lower bound of truncation, $\langle n_{max} \rangle$, in the Fock basis, can be obtained \cite{OCasta11}. It is built by taking the expectation values of $a^{\dagger}a$, and adding five times its quadratic deviation. In the superradiant phase it reads

\begin{equation}
\langle n_{max} \rangle 
=\mathcal{N}\gamma^{2}\left(1-\left(\frac{\sqrt{\omega\omega_{0}}}{2\gamma}\right)^{4}\right)+5\sqrt{\mathcal{N}\gamma^{2}\left(1-\left(\frac{\sqrt{\omega\omega_{0}}}{2\gamma}\right)^{4}\right)}.
\label{Cas}
\end{equation}
\begin{figure}[h!] 
\centering
\includegraphics[width=0.45 \textwidth]{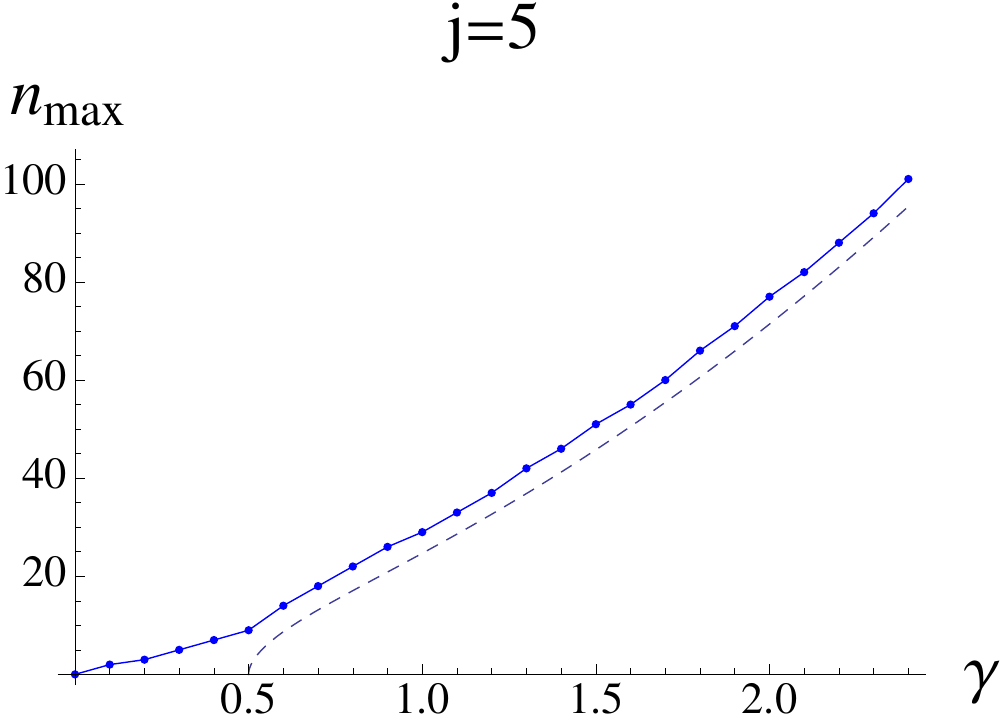}
\includegraphics[width=0.45 \textwidth]{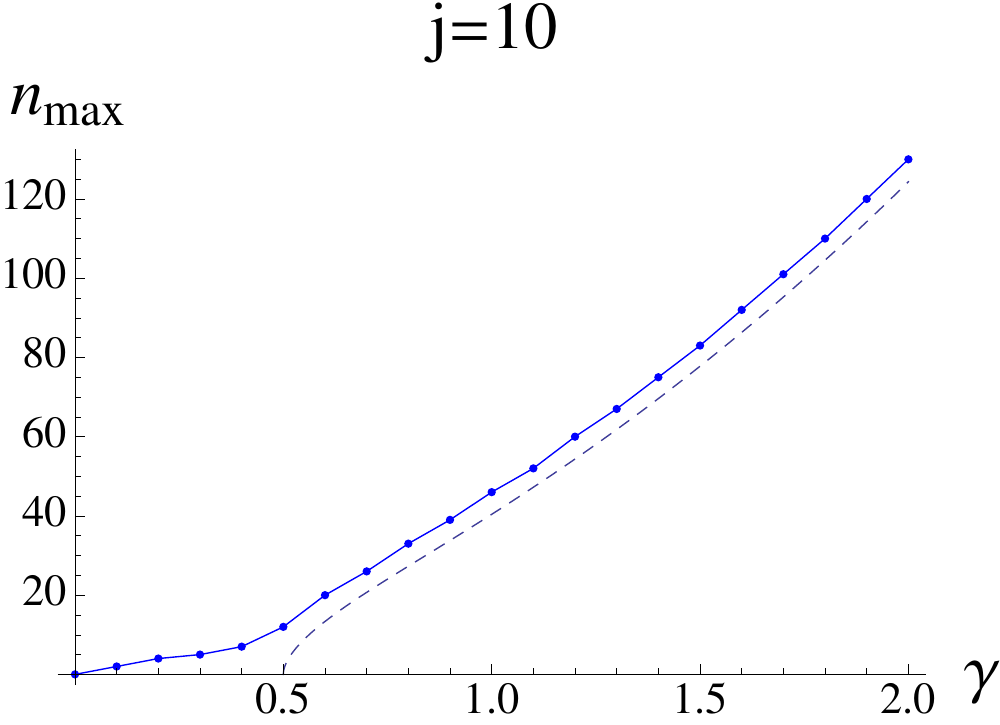}
\caption{Comparison between $n_{max}$ calculated via the $\Delta E$ criteria (dotted line) and Eq. (\ref{Cas}) (dashed line), for $j=5$ (left), and $j=10$ (right). Tolerance $\epsilon=1$x$10^{-6}$. }
\label{fig3}
\end{figure}

In figure \ref{fig3} we compare this expression with the $n_{max}$ obtained numerically using the $\Delta E$ convergence criteria. The agreement between the two curves is remarkably good. It points out that the $\Delta E$ criteria is enough to obtain the right value of the truncation. The above expression describes the minimal dimension of the photon sector necessary to calculate the ground state energy with the desired precision, showing that it grows linearly with the number of atoms ${\mathcal N}$, and quadratically with the interaction strength $\gamma$, as mentioned above. 

\subsection{Out of resonance}

\begin{figure}[h!] 
\centering
\includegraphics[width=0.45 \textwidth]{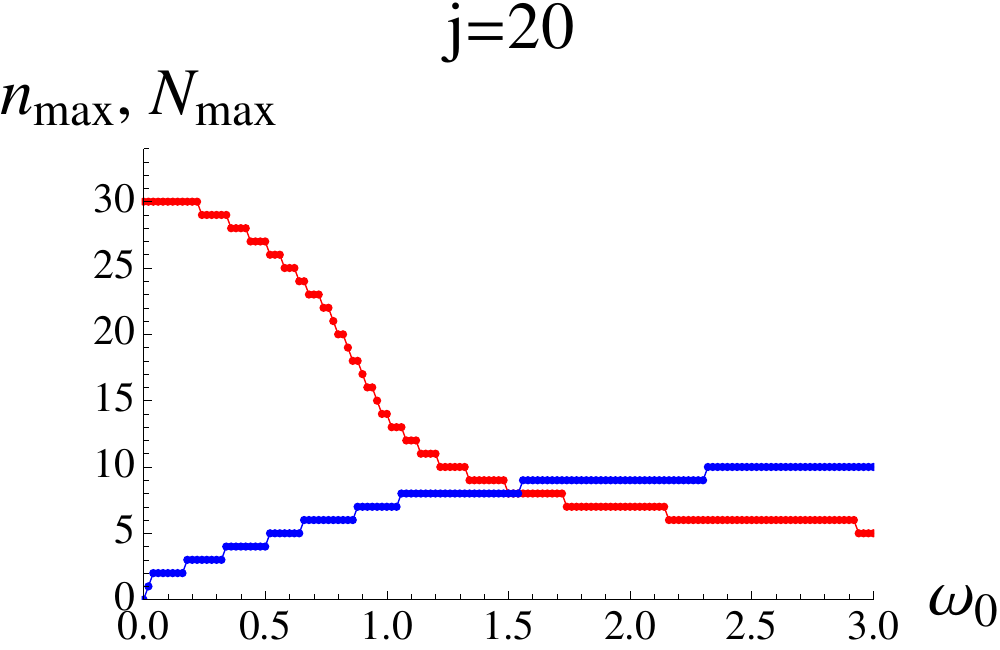}
\qquad
\includegraphics[width=0.45 \textwidth]{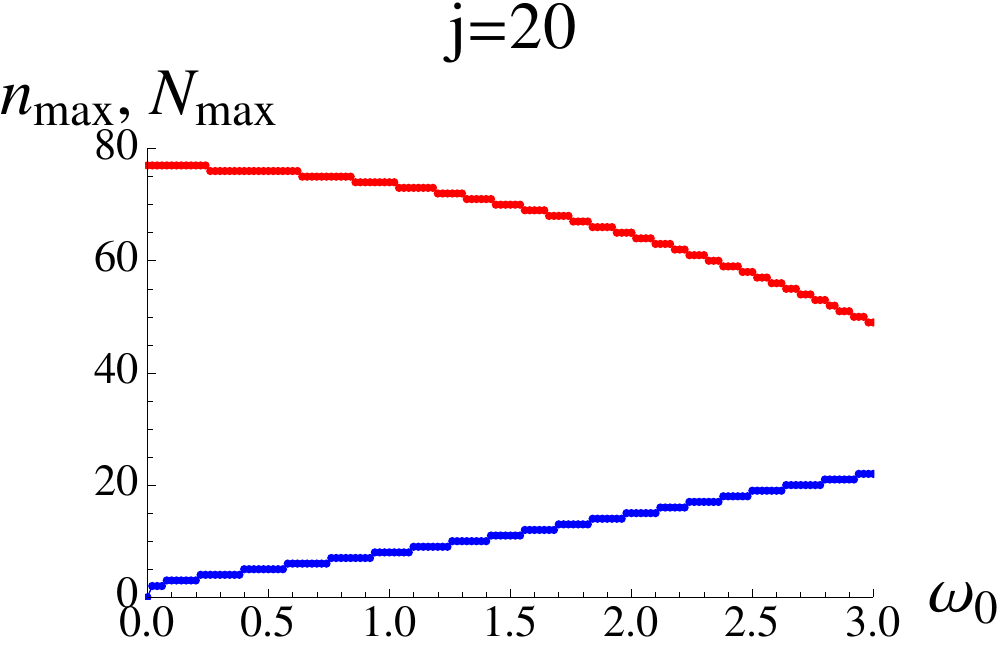}
\caption{$n_{max}$ (upper, red line) and $N_{max}$ (lower, blue line) as functions of $\omega_{0}$ with $\omega=1$ and $\gamma=0.5$ (left) and $\gamma=1.0$ (right) for $j=20$ . Tolerance $\epsilon=1$x$10^{-6}$.}
\label{fig4}
\end{figure}

In order to analyze the the truncation behavior out of resonance, we fix $\omega=1$ and vary the value of $\omega_{0}$ for $j= 20$ and two values of $\gamma$. In figures \ref{fig4} we show the results for $\gamma=0.5$ and $1.0$ respectively. In this case the normal region is defined by $\omega_0 > 4\, \omega \, \gamma^2$. It means that the superradiant region comprises  $\omega_0 < 1.0$ in Fig. \ref{fig4}, left, and $\omega_0 < 4.0$ in Fig. \ref{fig4}, right.

As was pointed out above, when $\omega_{0} \rightarrow 0$, for every $\gamma$ and $j$, the coherent basis is the best option to diagonalize the Hamiltonian because it is the exact solution. As $\omega_{0}$ increases, a crossing between the curves describing $n_{max}$ and $N_{max}$ as functions of $\omega_{0}$ can be observed. It takes place in the normal region, where $\gamma << \gamma_c$, i.e. the coupling constant is very small compared with the atomic excitation energy. Only in this region the Fock basis seems to require a truncation smaller than the coherent basis.

\subsection{Precision vs truncation}

\begin{figure}[h!] 
\centering
\includegraphics[width=0.45 \textwidth]{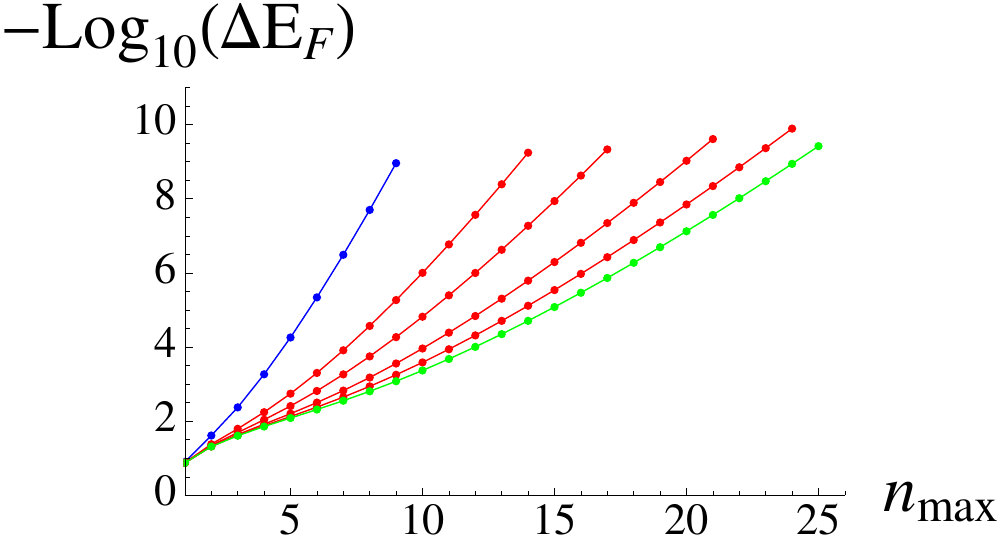}
\includegraphics[width=0.45 \textwidth]{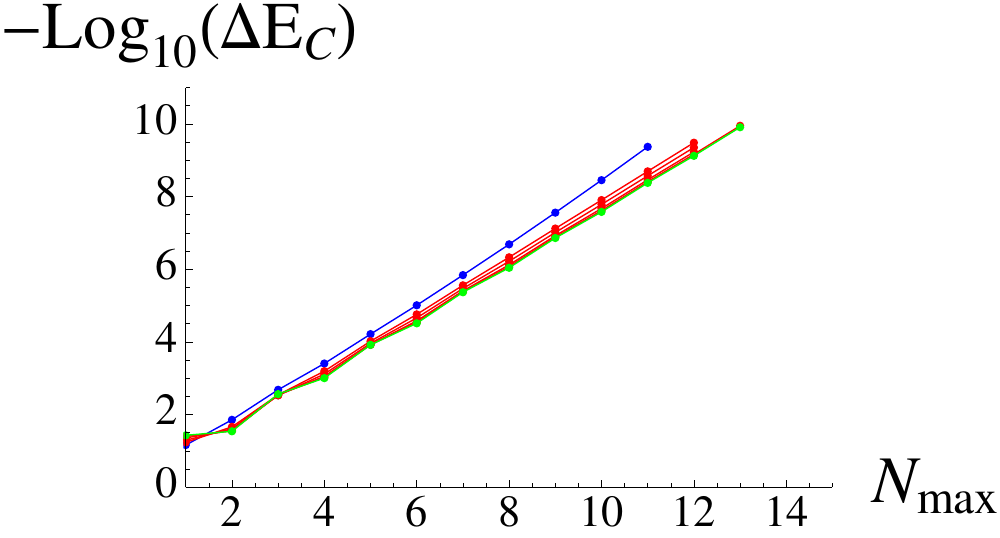}
\caption{(Color online). $\Delta E$ as function of $n_{max}$ (left) and $N_{max}$ (right). From left to right $j=1$ (blue), $5$, $10$, $20$, $30$ and $40$ (green). For $\gamma=0.5$ in resonance. }
\label{fig5}
\end{figure}

It is possible to select the desired precision in the ground state energy by knowing how fast $\Delta E$ goes to zero in both basis as $n_{max}$ and $N_{max}$ increase. For several $j$ representatives and with $\gamma=0.5$, in resonance, we show the results in figure \ref{fig5}. Working in this parameter region around the QPT region no preference is given to any of the two basis. Here $\Delta E_{F}$ and $\Delta E_{C}$ account for the Fock and coherent basis, Eq. (\ref{DeltaEF}).

As shown in figure \ref{fig5}, increasing the energy precision in the Fock basis demands a larger $n_{max}$, which grows also with $j$. In the coherent basis, left figure, the required values of $N_{max}$ are smaller, and seem to be quite independent of $j$ in the cases analyzed here. 

A linear fit for $j=40$ gives us the following relation between $N_{max}$ and $\Delta E_{C}$:
\begin{equation}
-Log_{10}\Delta E_{C}=0.278+ 0.732 N_{max} \Rightarrow\,\,\, \Delta E_{C}=0.526\,\,10^{-0.732 N_{max}}.
\end{equation}

\section{Conclusions}

To obtain the eigenvalues and eigenvectors of the Dicke Hamiltonian for a finite number of atoms it is necessary to perform a numerical diagonalization, employing a truncated boson number space. Two basis, associated with the two integrable limits of the Hamiltonian, are used along this work. The Fock basis corresponds to the zero interaction limit is the most common used, however, it consumes a lot of computing resources and becomes impractical to study the superradiant region for more than a few dozens of atoms. In the present article we have shown that, in most of the Hamiltonian's parameter regions including the QPT, the coherent basis requires a significative smaller truncation. In the accompanying work \cite{Hir13} we present a similar analysis for a convergence criteria based in the wave function, and for excited states.

We thank O. Casta\~nos, R. L\'opez-Pe\~na and E. Nahmad for many useful and interesting conversations.This work was partially supported by CONACyT-M\'exico  and PAPIIT-UNAM 102811.

\end{document}